# CMOL: SECOND LIFE FOR SILICON?

*Konstantin K. Likharev*

Stony Brook University, NY 11794-3800, U.S.A.

**ABSTRACT**

This report is a brief review of the recent work on architectures for the prospective hybrid CMOS/nanowire/nanodevice ("CMOL") circuits including digital memories, reconfigurable Boolean-logic circuits, and mixed-signal neuromorphic networks. The basic idea of CMOL circuits is to combine the advantages of CMOS technology (including its flexibility and high fabrication yield) with the extremely high potential density of molecular-scale two-terminal nanodevices. Relatively large critical dimensions of CMOS components and the "bottom-up" approach to nanodevice fabrication may keep CMOL fabrication costs at affordable level. At the same time, the density of active devices in CMOL circuits may be as high as $10^{12}$ cm$^2$ and that they may provide an unparalleled information processing performance, up to $10^{20}$ operations per cm$^2$ per second, at manageable power consumption.

## 1. INTRODUCTION

It is now generally accepted [1] that the current VLSI paradigm (based on a combination of lithographic patterning, CMOS circuits, and Boolean logic) can hardly be extended into a-few-nm region. The most fundamental reason is that at gate length below 10 nm, the sensitivity of parameters (most importantly, the gate voltage threshold) of silicon MOSFETs to inevitable fabrication spreads grows exponentially [2, 3]. As a result, the gate length should be controlled with a few-angstrom accuracy, far beyond even the long-term projections of the semiconductor industry [1]. (Similar problems are faced by the lithography-based single-electron devices [3].) Even if such accuracy could be technically implemented using a sophisticated patterning technology, this would send the fabrication facilities costs (growing exponentially even now) skyrocketing, and lead to the end of the Moore's Law some time during the next decade. This is why there is a rapidly growing consensus that the impending crisis of the Moore Law may be deferred only by a radical paradigm shift from the lithography-based fabrication to the "bottom-up" approach based on nanodevices with Nature-fixed size, e.g., specially designed molecules. Since the functionality of such nanodevices is relatively low [3], they almost necessarily should be used just as an add-on to a CMOS subsystem. Such combination allows the functionality, flexibility, and reliability of CMOS circuits (and all the enormous infrastructure created by the electronic industry for their design and fabrication) to be used in full extent.

Several proposals of such hybrid CMOS/nanodevice circuits were put forward recently (for their reviews, see Refs. 4, 5). Essentially, all of these proposals are based on the use of two-terminal nanodevices with the functionality illustrated by Fig. 1a. At low voltages, such devices may behave as usual diodes, but the application of a higher voltage may switch them between low-resistive (ON) and high-resistive (OFF) states. (This means that the device incorporates single-bit internal memory.) Such "programmable diode" functionality may be achieved in several ways, for example, by switching between two atomic configurations of a molecule –see, e.g., Refs. 6-8. However, nanosecond-scale operation speed requires electron switches, e.g., single-electron latches (Fig. 1b) [9] whose low-temperature prototypes have already been demonstrated [10].

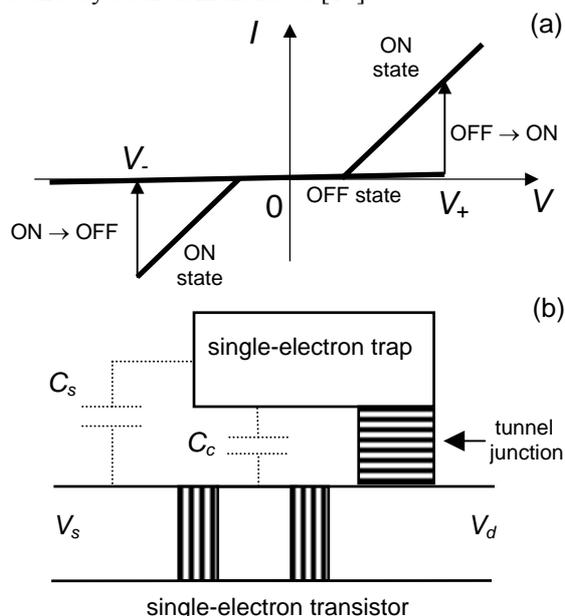

Fig. 1. Programmable diode ("latching switch"): (a) I-V curve (schematically) and (b) possible implementation using single-electron devices.

Room-temperature versions of such devices may be readily implemented using relatively simple organic molecules [11]. (Their main components, molecular single-electron transistors, have already been demonstrated by several research groups [12-16].)

The decisive advantage of using two-terminal nanodevices like programmable diodes is that reproducible fabrication of their molecular versions seems rather feasible – see, e.g., the recent experiments [17].

## 2. CMOL

Our group is working on a particular type of hybrid CMOS/nanodevice circuits, dubbed CMOL, which look more plausible for implementation [3, 18]. As in several earlier proposals [4-8], nanodevices in CMOL circuits are formed (e.g., self-assembled) at each crosspoint of a "crossbar" array, consisting of two levels of nanowires (Fig. 2). However, in order to overcome the CMOS/nanodevice interface problems pertinent to earlier proposals, in CMOL circuits the interface is provided by pins that are distributed all over the circuit area, on the top of the CMOS stack. (Silicon-based technology necessary for fabrication of pins with nanometer-scale tips has been already developed in the context of field-emission arrays [19].) As Fig. 2c shows, pins of each type (reaching to either the lower or the upper nanowire level) are arranged into a square array with side $2\beta F_{CMOS}$, where $F_{CMOS}$ is the half-pitch of the CMOS subsystem, and $\beta$ is a dimensionless factor larger than 1 that depends on the CMOS cell complexity. The nanowire crossbar is turned by angle $\alpha = \arcsin(F_{nano}/\beta F_{CMOS})$ relative to the CMOS pin array, where $F_{nano}$ is the nanowiring half-pitch.

By activating two pairs of perpendicular CMOS lines, two pins (and two nanowires they contact) may be connected to CMOS data lines (Fig. 2b). As Fig. 2c illustrates, this approach allows a unique access to any nanodevice, even if $F_{nano} \ll F_{CMOS}$ - see Ref. 18 for a detailed discussion of this point. If the nanodevices have a sharp current threshold, like the usual diodes, such access allow to test each of them. Moreover, if the nanodevice may be switched between two internal states (e.g. is a programmable diode), switching between the states may be achieved by applying voltages $\pm V_W$ to the selected nanowires, so that voltage $V = \pm 2V_W$ applied to the selected nanodevice exceeds the corresponding switching threshold, while half-selected devices (with $V = \pm V_W$) are not disturbed.

Two advantages of CMOL circuits over other crossbar-type hybrids look most important. First, due to the uniformity of the nanowiring/nanodevice levels of CMOL, they do not need to be precisely aligned with each other and the underlying CMOS stack, thus allowing the use for nanowire formation of advanced patterning techniques [20], [21] which lack precise layer alignment. Second, CMOL circuits may work with two-terminal nanodevices (e.g., single-electron latching switches). Still, CMOL, similarly to all other nanodevice-based technologies, requires defect-tolerant circuit architectures, since the fabrication yield of such devices will hardly ever approach 100% as closely as that achieved for the semiconductor transistors.

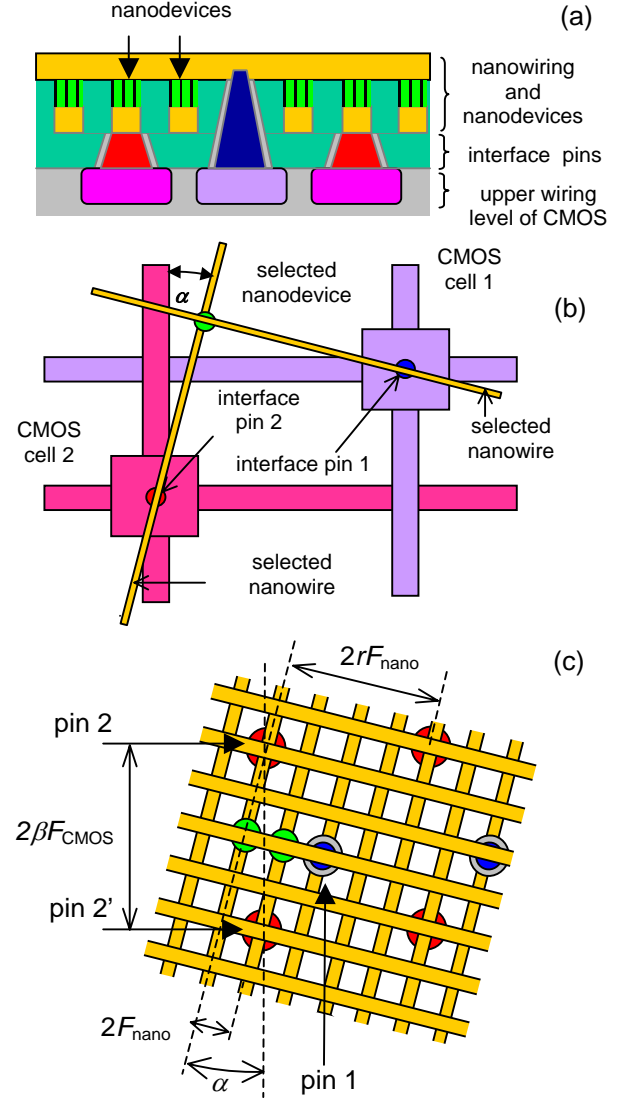

Fig. 2. Low-level structure of the generic CMOL circuit: (a) schematic side view; (b) the idea of addressing a particular nanodevice, and (c) zoom-in on several adjacent interface pins to show that any nanodevice may be addressed via the appropriate pin pair (e.g, pins 1 and 2 for the leftmost of the two shown devices, and pins 1 and 2' for the rightmost device). On panel (b), only the activated CMOS lines and nanowires are shown, while panel (c) shows only two devices. (In reality, similar nanodevices are formed at all nanowire crosspoints.)

## 3. MEMORIES

The most straightforward potential application of CMOL circuits are embedded memories and stand-alone memory chips, with their simple matrix structure. In such memories, each nanodevice would play the role of a single-bit memory cell, while the CMOS subsystem may be used for coding, decoding, line driving, sensing, and input/output functions.

We have carried out [22, 23] a detailed analysis of CMOL memories with global and quasi-local ("dash") structure of matrix blocks, including the application of two major techniques for increasing their defect tolerance: the memory matrix reconfiguration (the replacement of several rows and columns, with the largest number of bad memory cells, for spare lines), and advanced error correction codes. Figure 3 shows the final result of that analysis: the optimized total chip area per useful bit, as a function of the nanodevice yield.

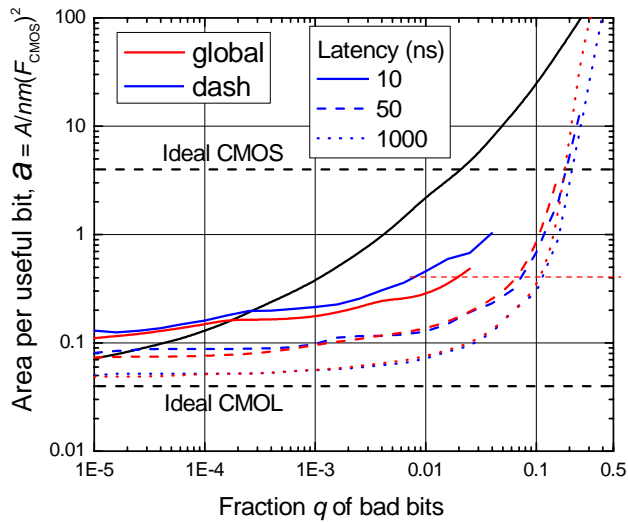

Fig. 3. The optimized area per useful bit as a function of single nanodevice yield, for two different CMOL memories (with "dash" and global block structure) and for several values of access time.

The results show that the memories may be rather defect-tolerant: a 90% yield may be achieved with 8 to 10% of bad nanodevices (depending on the required access time). When such yield has been achieved, the performance of CMOL memories will be extremely impressive. For example, the normalized cell area $a \equiv A/N(F_{CMOS})^2 = 0.4$ (red dashed line on Fig.3) at $F_{CMOS} = 32$ nm means that a memory chip of a reasonable size (2 × 2 cm$^2$) can store about 1 terabit of data - crudely, one hundred Encyclopedia Britannica's.

## 4. DIGITAL LOGIC CIRCUITS

The reconfiguration is the most efficient technique for coping with defective nanodevices in hybrid circuits. This is why the most significant published proposals for the implementation of logic circuits using CMOL-like hybrid structures had been based on reconfigurable regular structures like the field-programmable gate arrays (FPGA). Before our recent work, two FPGA varieties had been analyzed, one based on look-up tables (LUT) and another one using programmable-logic arrays (PLA). Unfortunately, all these approaches run into substantial problems – see Ref. 18 for discussion.

Recently, D. Strukov suggested [24] an alternative approach to Boolean logic circuits based on CMOL concept, that is close to the so-called cell-based FPGA [25]. In this approach an elementary CMOS cell includes two pass transistors and an inverter, and is connected to the nanowire/nanodevice subsystem via two pins (Fig. 4a). Disabling the CMOS inverters allows to carry out the circuit reconfiguration via cell's pass transistors. On the other hand, enabling the inverter turns the cell into a NOR gate (Fig. 4b), generally with an almost arbitrary fan-in.

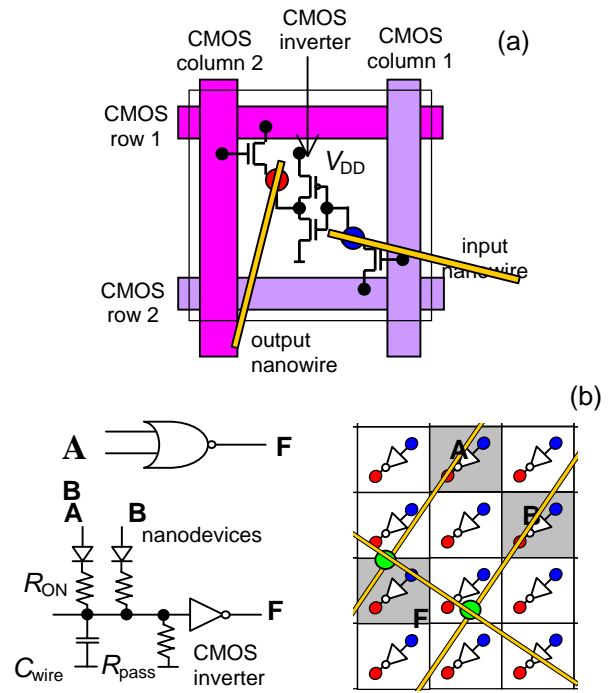

Fig. 4. CMOL FPGA: (a) logic cell schematics, and (b) implementation of a fan-in-two NOR gate.

First results for CMOL FPGA have been obtained using a simple, two-step approach to reconfiguration, in which the desired circuit is first mapped on the apparently perfect (defect-free) CMOL fabric, and then is reconfigured around defective components using a simple algorithm [24]. The Monte Carlo simulation (so far only for the "stack-on-open"-type defects which are expected to dominate in CMOL circuits) has shown that even this simple configuration procedure may ensure very high defect tolerance. For example, the reconfiguration of a 32-bit Kogge-Stone adder, mapped on the CMOL fabric with realistic values of parameters, may allow to achieve the 99% circuit yield (sufficient for a ~90% yield of properly organized VLSI chips), with as may as 22% of defective devices, while the defect tolerance of another key circuit, a fully-connected 64-bit crossbar switch, is about 25%.

Our most striking result was that such high defect tolerance may coexist with high density and performance, at acceptable power consumption. For example, our estimates have shown [24] that for the total power of 200 W/cm$^2$ (planned by the ITRS for the long-term CMOS technology nodes), an optimization may bring the logic delay of the 32-bit Kogge-Stone adder down to just 1.9 ns, at the total area of 110 μm$^2$, i.e. provide an area-delay product of 150 ns-μm$^2$, for realistic values $F_{CMOS}$ = 32 nm and $F_{nano}$ = 8 nm (Fig. 5). A minor error in that work (which was found later [26]) has led to underestimation of the actual delay by a factor close to 2.5. Still, the area-delay product compares very favorably with the estimated 70,000 ns-μm$^2$ (with 1.7 ns delay and 39,000 μm$^2$ area) for a fully CMOS FPGA implementation of the same circuit (with the same $F_{CMOS}$).

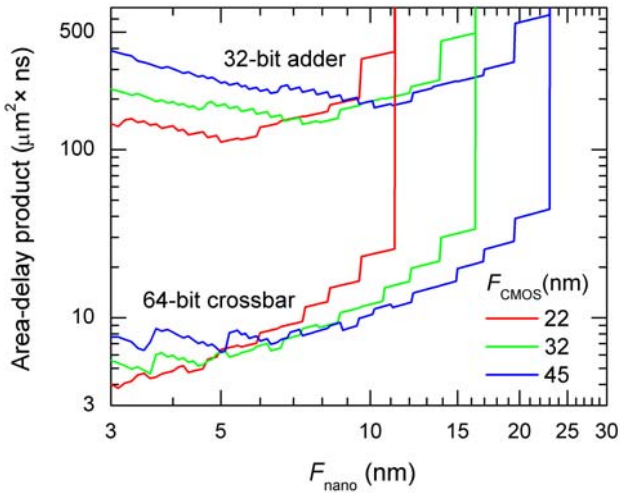

Fig. 5. CMOL FPGA performance as a function of nanowire half-pitch $F_{nano}$ (at the total power fixed at 200 W/cm$^2$) for two digital circuits, each for three long-term CMOS technology nodes [24].

Very recently, these calculations were extended [26] to all 20 circuits of the so-called Toronto benchmark set [27]. In order to accomplish this task, the first CMOL CAD tool has been developed using a combination of known CMOS FPGA design algorithms such as SIS, T-VPack and VPR [28, 29] with a CMOL-specific, custom routing program. There is still a lot of place for improvement in this tool, so that the preliminary results for area and (especially) logic delay for CMOL FPGA, presented in the right part of Table 1, should be considered just as upper bounds. Still, even these preliminary results show almost a spectacular density advantage (on the average, about two orders of magnitude) over the purely CMOS circuits and a considerable leading edge over a competing (and less feasible) hybrid circuit concept, so-called nanoPLA [31].

TABLE 1. Results of performance calculations for circuits of the Toronto 20 benchmark set implemented using CMOS and CMOL FPGAs, for $F_{CMOS}$ = 45 nm and $F_{nano}$ = 4.5 nm. Delay results for CMOL are just an upper bound rather than the critical path latency.

| Circuit | CMOS FPGA | | CMOL FPGA | | | |
|---|---|---|---|---|---|---|
| | Area (μm$^2$) | Delay (ns) | CMOS cells | Nano-devices | Area (μm$^2$) | Delay (ns) |
| Alu4 | 137700 | 5.1 | 1854 | 9788 | 1004 | 18.4 |
| apex2 | 166050 | 6.0 | 1928 | 11365 | 914 | 20.7 |
| apex4 | 414619 | 5.5 | 1176 | 7781 | 672 | 15.2 |
| bigkey | 193388 | 3.1 | 2065 | 10207 | 829 | 16.0 |
| clma | 623194 | 13.1 | 7585 | 48746 | 9308 | 59.9 |
| des | 148331 | 4.2 | 2321 | 12610 | 1097 | 22.3 |
| diffeq | 100238 | 6.0 | 2004 | 10799 | 1194 | 58.3 |
| dsip | 148331 | 3.2 | 1615 | 9905 | 829 | 20.7 |
| elliptic | 213638 | 8.6 | 4799 | 25415 | 4581 | 64.6 |
| ex1010 | 391331 | 9.0 | 2986 | 28746 | 3486 | 34.3 |
| ex5p | 100238 | 5.1 | 902 | 6875 | 829 | 21.5 |
| frisc | 230850 | 11.3 | 4715 | 25869 | 4199 | 91.0 |
| misex3 | 124538 | 5.3 | 1397 | 9211 | 1004 | 19.2 |
| pdc | 369056 | 9.6 | 4752 | 14841 | 4979 | 43.1 |
| s298 | 166050 | 10.7 | 1030 | 10161 | 829 | 35.9 |
| s38417 | 462713 | 7.3 | 8289 | 53156 | 9308 | 41.5 |
| s38584 | 438413 | 4.8 | 6502 | 50275 | 9872 | 51.1 |
| seq | 151369 | 5.4 | 1832 | 11027 | 1296 | 18.4 |

| spla | 326025 | 7.3 | 4240 | 24808 | 2994 | 31.9 |
| tseng | 78469 | 6.3 | 1866 | 4918 | 1194 | 59.9 |

## 5. MIXED-SIGNAL NEUROMORPHIC CIRCUITS

One more possible application of CMOL circuits is neuromorphic networks (see, e.g., Ref. 32). We have explored a specific architecture of such networks, called Distributed Crossbar Networks ("CrossNets") [11, 34], which are uniquely suitable for CMOL implementation. In each CrossNet (Fig. 6), relatively sparse neural cell bodies ("somas") are implemented in the CMOS subsystem, while the much denser latching switches are used as elementary synapses. The mutually perpendicular nanowires of the CMOL crossbar naturally implement the axons and dendrites which carry signals between the cells, allowing one cell to be directly connected to a virtually unlimited number $M$ of other cells. Due to this parallelism, CrossNets can be spectacularly resilient, with visible performance degradation at as many as ~50% of bad nanodevices [34].

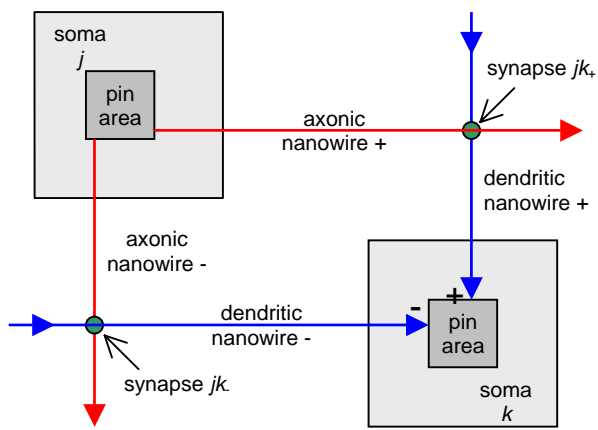

Fig. 6. The simplest feedforward CrossNet.

CrossNet functionality is strongly dependent on the distribution of the somas over the axon/dendrite/synapse field (Fig. 7). For example, "FlossBars" (Fig. 7a) have a layered topology and hence can be used to implement multi-layer perceptron (MLPs), while "InBars" (Fig. 7b) have an "interleaved" structure which is natural for the implementation of recurrent networks – see, e.g., Ref. 32.

CrossNet training faced several challenges including the binary character of the elementary synapse (latching switch) and a certain statistical uncertainty of its switching. In our recent work [34] we have proved that, despite these limitations, CrossNets can be taught, by at least two different methods, to perform virtually all the major functions demonstrated earlier with usual neural networks, including the corrupted pattern restoration in the recurrent quasi-Hopfield mode and pattern classification in the feedforward MLP mode.

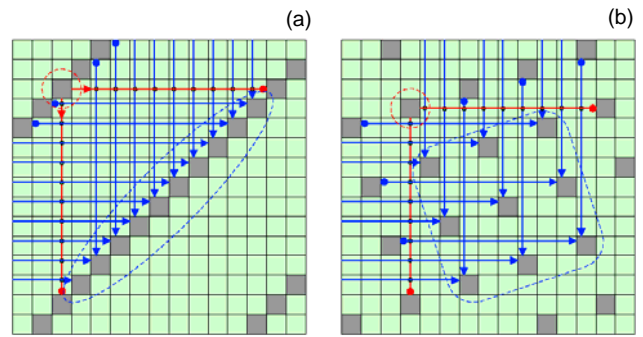

Fig. 7. Two particular CrossNet species: (a) FlossBar (shown for the connectivity parameter $M$ equal to 10) and (b) InBar (for $M = 9$). For clarity, the figures show only the axons, dendrites, and synapses providing connections between one soma (indicated by the dashed red circle) and its direct signal recipients (inside the dashed blue lines), for the simplest (non-Hebbian feedforward) networks.

The importance of this result is in the CrossNet's potential unparalleled density and speed [11, 34]: for realistic parameters, the cell density may exceed that of cerebral cortex (above $10^7$ cells per cm$^2$), while the average cell-to-cell communication delay may be as low as ~10 ns (i.e., about six orders of magnitude lower than that in the brain), at acceptable power. Even putting aside the exciting long-term prospects of creating high-speed artificial brain-like systems [34], CMOL CrossNet chips of modest size might be used for important present-day problems, e.g., online recognition of a person in a large crowd [35].

## 6. CONCLUSIONS

I believe there is a chance for the development, within perhaps the next 10 to 15 years and maybe substantially earlier, of hybrid "CMOL" integrated circuits that will allow to extend Moore's Law to the few-nm range. Estimates show that such circuits could be used for several important applications, notably including terabit-scale memories, reconfigurable digital circuits with multi-teraflops-scale performance, and mixed-signal neuromorphic networks that may, for the first time, compete with biological neural systems in areal density, far exceeding them in speed, at acceptable power dissipation.

The major challenges on the way toward practical CMOL circuits include the development of high-yield techniques for formation (e.g., chemically-directed molecular self-assembly) of single-electron latching switches, even better architectures for digital CMOL circuits, and training mixed-signal neuromorphic networks to perform more advanced, eventually intellectual information processing tasks.

## ACKNOWLEDGMENTS

Useful discussions of the issues considered in this paper with P. Adams, J. Barhen, W. Chen, S. Das, J. Ellenbogen, D. Hammerstrom, R. Karri. P. Kuekes, J. H. Lee, X. Liu, J. Lukens, X. Ma, A. Mayr, V. Protopopescu, M. Reed, M. Stan, D. Strukov, Ö. Türel, and S. R. Williams are gratefully acknowledged. The work has been supported in part by AFOSR, ARDA, NSF, and MARCO via FENA Center. This paper is an updated and extended version of the invited talk [36].